\documentclass[10pt,letterpaper]{article}
\usepackage[top=0.85in,left=2.75in,footskip=0.75in]{geometry}

\usepackage{amsmath,amssymb}

\usepackage{changepage}

\usepackage[utf8x]{inputenc}

\usepackage{textcomp,marvosym}

\usepackage{cite}

\usepackage[right]{lineno}

\usepackage{microtype}
\DisableLigatures[f]{encoding = *, family = * }

\usepackage[table]{xcolor}

\usepackage{array}

\usepackage{stackengine}
\usepackage{mathtools}
\usepackage{floatrow}
\usepackage{graphicx}
\usepackage{tabularx}
\usepackage{multirow}
\usepackage{wrapfig}
\usepackage{float}
\usepackage[font=footnotesize,labelfont=bf]{caption}
\usepackage{subcaption}
\usepackage{todonotes}

\graphicspath{{./figures_FINAL/}}

\newcommand{\ddt}[1]{\ensuremath{\frac{d #1}{dt}}}

\newcommand*{\pd}[3][]{\ensuremath{\frac{\partial^{#1} #2}{\partial #3}}}
\newcommand {\rzero}{\mathcal{R}_0}
\newcommand {\dt}{\Delta t}
\newcommand {\dx} {\Delta x}
\newcommand{\ntot}{\ensuremath{N}}
\newcommand{\nabm}{\ensuremath{N}}

\newcolumntype{+}{!{\vrule width 2pt}}

\newlength\savedwidth

\raggedright
\usepackage[aboveskip=1pt,labelfont=bf,labelsep=period,justification=raggedright,singlelinecheck=off]{caption}

\makeatletter
\renewcommand{\@biblabel}[1]{\quad#1.}
\makeatother

\usepackage{lastpage,fancyhdr,graphicx}
\usepackage{epstopdf}
\fancyheadoffset[L]{2.25in}
\fancyfootoffset[L]{2.25in}
\begin{document}
\vspace*{0.2in}

\begin{flushleft}
{\Large
\textbf\newline{Choice of spatial discretisation influences the progression of viral infection within multicellular tissues} 
}
\newline
\\
Thomas Williams\textsuperscript{1},
James McCaw\textsuperscript{1,2},
James Osborne*\textsuperscript{1}
\\
\bigskip
\textbf{1} School of Mathematics and Statistics, University of Melbourne, Australia
\\
\textbf{2} Centre for Epidemiology and Biostatistics, Melbourne School of Population and Global Health, University of Melbourne, Australia
\\

\bigskip

%
%





* jmosborne@unimelb.edu.au

\end{flushleft}
\section*{Abstract}
There has been an increasing recognition of the utility of models of the spatial dynamics of viral spread within tissues. Multicellular models, where cells are represented as discrete regions of space coupled to a virus density surface, are a popular approach to capture these dynamics. Conventionally, such models are simulated by discretising the viral surface and depending on the rate of viral diffusion and other considerations, a finer or coarser discretisation may be used. The impact that this choice may have on the behaviour of the system has not been studied. Here we demonstrate that, if rates of viral diffusion are small, then the choice of spatial discretisation of the viral surface can have quantitative and even qualitative influence on model outputs. We investigate in detail the mechanisms driving these phenomena and discuss the constraints on the design and implementation of multicellular viral dynamics models for different parameter configurations.

\section*{Author summary}
Here we analyse the impact that the resolution of spatial discretisation has on the behaviour of spatial models of viral dynamics. We show that the extent of this influence depends on the size of the diffusion coefficient, and that, if the diffusion coefficient is sufficiently small - for example, small enough to describe tight plaque-forming dynamics - the choice of discretisation of the viral surface relative to the cell grid can appreciably change key predictions of the model. We show that the resolution of spatial discretisation modulates the timescale of infection, and identify the mechanisms by which this occurs. This work provides theory to inform the development of multicellular models of viral dynamics, which is in short supply in the literature.

\section*{Introduction}
Mathematical modelling provides a useful toolkit for studying the spread of viral infection within tissues, both \emph{in vivo} and \emph{in vitro}. Over a number of decades a substantial literature has evolved, driving significant progress in our understanding of the mechanisms that drive the virological and immunological dynamics of a number of viruses, including HIV \cite{perelson_hiv_review, graw_and_perelson_spatial_hiv}, hepatitis C \cite{guedj_et_al_hcv_daclatasvir, neumann_et_al_hcv_interferon}, SARS-CoV-2 \cite{sego_et_al_covid_model, getz_rapid_community_driven_covid_model, hernandez_vargas_modelling_covid_in_humans}, influenza \cite{beauchemin_and_handel_influenza_review, baccam_et_al_influenza_kinetics},  and oncolytic viruses \cite{rodriguez_brenes_et_al_virus_spread_from_low_moi, paiva_et_al_multiscale_oncolytic_virotherapy}. The importance of this work has recently been made apparent by the ongoing COVID-19 pandemic. The outbreak has prompted a large output of modelling work which has aided in the development of antiviral therapies, and an understanding of disease pathogenesis \cite{goyal_et_al_potency_and_timing_covid_antivirals, perelson_and_ke_covid_review, jenner_et_al_covid_virtual_patient_cohort}.

An emerging trend in the field of viral dynamics is the development of spatially-resolved models (e.g., \cite{graw_and_perelson_spatial_hiv, sego_et_al_covid_model, getz_rapid_community_driven_covid_model, wodarz_et_al_complex_spatial_dynamics_oncolytic, paiva_et_al_multiscale_oncolytic_virotherapy}). Unlike classical models for within host viral spread, which assume mean-field dynamics, usually described by ordinary differential equations (ODEs), spatial models describe the spatial distribution of infection and track the spreading front of viral invasion. Spatial models provide a more detailed account of infection and its interaction with the immune response and the predicted dynamics can often differ substantially from those obtained from comparable ODE models \cite{holder_et_al_design_considerations_influenza, beauchemin_and_handel_influenza_review}. Furthermore, depending on availability of suitable spatially-resolved data to support inference, estimated parameters are also often significantly different to those estimated from non-spatial models \cite{holder_et_al_design_considerations_influenza, graw_and_perelson_spatial_hiv}.

There are several ways of implementing spatial structure in a viral dynamics model, for example, using neighbour-based, cellular automata methods where cells are rendered explicitly (e.g., \cite{funk_et_al_spatial_models_virus_immune, beauchemin_well_mixed_assumption, kumberger_et_al_accounting_for_space_cell_to_cell, rodriguez_brenes_et_al_virus_spread_from_low_moi}), or spatially-continuous partial differential equations (PDEs) (e.g., \cite{bocharov_et_al_dynamics_virus_infection_space_time, bocharov_et_al_spatiotemporal_dynamics_virus, quirouette_et_al_influenza_localisation_model}). An increasingly popular choice of modelling structure combines the two approaches \cite{sego_et_al_cellularisation, sego_et_al_covid_model, getz_rapid_community_driven_covid_model, whitman_et_al_spatiotemporal_host_virus_influeza, levin_et_al_T_cell_search_influenza}. These multicellular models represent cells as a grid of discrete regions of space, whilst virus density is described by a PDE. Usually, virus is produced by cells in the region of the cell, and then spreads across the cell grid according to linear diffusion. This approach is based on the assumption of different spatial scales of cells and viruses, where the size of a virus is assumed to be negligible compared to the size of a cell \cite{sego_et_al_cellularisation, sego_et_al_covid_model}.

However, when simulating such models numerically, the three spatial scales which are present --- virus, cells, and the size of the tissue or computational domain --- introduce a layer of complexity when it comes to the discretisation of space. Whilst some works have provided some discussion of practical concerns in the design of spatial models of this nature, and in particular the role of the viral diffusion coefficient \cite{holder_et_al_design_considerations_influenza, sego_et_al_cellularisation, gallagher_et_al_causes_and_consequences}, there has been no direct analysis to our knowledge of the role of the resolution of spatial discretisation in these models. However, the choice of $\dx$, the spacing between discrete nodes or elements for the approximate viral surface, relative to the spacing of the cell grid, is non-trivial. At extremely fine refinement, where $\dx$ is much smaller than a cell diameter, errors may be kept very low, but computational costs may be very large, and moreover the dependence of the model on the exact arrangement of cells is very strong. On the other hand, if $\dx$ is too large, perhaps much larger than a cell diameter, some of this spatial detail may be lost and the outcome of simulations may be affected. It is not clear \emph{a priori} at what point, or under which conditions, model predictions start to become unreliable, or what behaviour is affected by the choice of spatial discretisation. 

The influence that this choice of resolution of spatial discretisation has on model simulations has been investigated in other domains of mathematical biology. In cardiac tissue electrophysiology, for example, one review studied the convergence of a range of published models under various conditions and found that results were highly dependent on the spatial discretisation \cite{niederer_et_al_cardiac_tissue_review}. Another work studied mechanisms causing spiral electrical wave breakup in cardiac tissue, and found that such behaviour could arise from discrete effects: either coarsening the resolution of the spatial discretisation, or modelling a reduction in cell coupling \cite{fenton_et_al_spiral_wave_breakup}. This finding suggests that, left unanalysed, artefacts arising from spatial discretisation of models could mask biologically important model outcomes. The above models of cardiac electrophysiology are based on continuous sheets of tissues \cite{niederer_et_al_cardiac_tissue_review, fenton_et_al_spiral_wave_breakup, ten_tusscher_et_al_model_human_ventricular_tissue}. This ignores the discrete qualities of individual cells, which is clearly a key feature of multicellular models of viral dynamics outlined above. However, to our knowledge, a detailed discussion of the role of spatial discretisation in the latter modelling framework has not been carried out. Here we conduct a number of experiments and simulations using a simple but typical multicellular model of viral dynamics, and show the influence of changing $\dx$ on a wide range of model outcomes under a range of different conditions.

\section*{Methods}

\subsection* {The TIV model and its agent-based form}
The classical model of virus dynamics within the host is the so-called TIV model, a simple but extremely robust ODE model of viral dynamics derived from mass-action kinetics \cite{lavigne_et_al_interferon_signalling_ring_vaccination}. The model has been well studied and applied to a wide variety of viral infections, including HIV (\cite{perelson_hiv_review} provides a review), Hepatitis C virus (e.g., \cite{neumann_et_al_hcv_interferon}), influenza (e.g., \cite{beauchemin_and_handel_influenza_review, baccam_et_al_influenza_kinetics}), and SARS-CoV-2 (e.g., \cite{hernandez_vargas_modelling_covid_in_humans, du_and_yuan_innate_and_adaptive_immune}), and yielded useful estimations of biological parameters in each case. We assume a population of target cells $T$ and infected cells $I$, as well as a global measure of extracellular viral load $V$. Target cells are assumed to become infected by virus at a rate $\beta$, infected cells are assumed to die at a rate $\delta$, and virus is produced by infected cells at rate $p$, and is cleared at rate $c$. These assumptions are deceptively powerful, since, for example, infected cell and virus clearance rates can implicitly account for effects of immune clearance. Some presentations of the TIV model include birth and death phenomena of target cells, however, here we will assume such effects are negligible, which is common for models of acute infections \cite{smith_and_perelson_influenza_review}. The model under these assumptions has the following form:

\begin{align} 
\label{eq:TIV_T}
\ddt{}\left(\frac{T}{\ntot}\right) &= -\beta \frac{T}{\ntot}V,\\
\label{eq:TIV_I}
\ddt{}\left(\frac{I}{\ntot}\right) &= \beta \frac{T}{\ntot}V - \delta \frac{I}{\ntot},\\
\label{eq:TIV_V}
\ddt{V} &= pI - cV.
\end{align}

\noindent Here $\ntot = T_0 + I_0$, where $T_0$ is the initial number of target cells and $I_0$ is the initial number of infected cells. Typically $I_0$ is small (or zero) and thus $\ntot \approx T_0$. Including dead cells, $\ntot$ represents the total number of cells in the model. As such, in this form, the model tracks the proportion of the cell population comprised of target cells and infected cells over time. Analysis has shown that much of the behaviour of the model depends on the basic reproduction number $\rzero = \beta p T_0 / (\delta c)$ (assuming $I_0 / \ntot \approx 0$) \cite{lavigne_et_al_interferon_signalling_ring_vaccination, beauchemin_and_handel_influenza_review}. $\rzero$ defines the number of infected cells produced by a single infected cell during its lifetime \cite{diekmann_next_generation_matrices}. As such, when $\rzero<1$, the infection will die out, whereas if $\rzero>1$ an infection can be established \cite{baccam_et_al_influenza_kinetics}.

Since we are interested in finding a spatially explicit multicellular equivalent of Eqs \eqref{eq:TIV_T}--\eqref{eq:TIV_V}, we now define cells as explicit, individual agents which occupy regions of physical space, and can be in one of three states: uninfected (T), infected (I), or dead (D). We associate with each cell, living or dead, an index $i \in \left\lbrace 1, 2, ... ~\nabm \right\rbrace$. We assume that the cells are arranged in a regular grid which tessellates the spatial domain of study, $\Omega$, resulting in a confluent monolayer of cells similar to an \emph{in vitro} assay. Here we assume cells are simply arranged in a rectangular grid, where each cell is assumed to be a unit square. It is worth mentioning, however, that a more precise formulation would be to assume a hexagonal grid, which has been applied in a number of other, agent-based models of viral infections and more accurately represents the packing of epithelial cells in a confluent monolayer \cite{whitman_et_al_spatiotemporal_host_virus_influeza, holder_et_al_design_considerations_influenza, kumberger_et_al_accounting_for_space_cell_to_cell, blahut_et_al_hepatitis_c_two_modes_of_spread}. For the purposes of our investigation of numerical discretisation, it will be sufficient and markedly more intuitive to use a rectangular cell grid, although we have also repeated our experiments for hexagonal cell grid with qualitatively the same results (figures not shown for sake of brevity). We do not anticipate that our results will be qualitatively any different for other grid configurations.

We assume that virus is produced uniformly over the domain of  infected cells, and obeys linear diffusion in the computational domain. As such, the virus equation, Eq \eqref{eq:TIV_V}, becomes the PDE

\begin{equation} \label{eq:virus_pde}
\pd{v}{t} = p\sum_{i \in \mathcal{I}} \frac{ \mathcal{X}_{S_i} (\mathbf{x})}{\left| S_i \right|} - cv +D \nabla^2 v,
\end{equation}

\noindent where we write $\mathcal{I}$ for the set of infected cells, $S_i$ for the region of space occupied by cell $i$, and $D$ for the diffusion coefficient. $\mathcal{X}_S (\mathbf{x})$ is the characteristic function

\begin{equation*}
\mathcal{X}_{S_i} (\mathbf{x}) = \begin{cases}
1, \qquad \mathbf{x} \in S_i,\\
0, \qquad \mathbf{x} \notin S_i.
\end{cases}
\end{equation*}

\noindent It is straightforward to verify that integrating Eq \eqref{eq:virus_pde} over $\Omega$, subject to appropriate boundary conditions, recovers the ODE form of the virus equation.

To replace the target cell and infected cell equations, Eqs \eqref{eq:TIV_T} and \eqref{eq:TIV_I}, with update rules for individual cells, we consider the Poisson process of the number of state transitions in a given time interval $t$. We consider the simple case of the transition from infected cell to dead cell, $I\rightarrow D$. Infected cell death occurs at rate $\delta$, so for a given infected cell $i$, the probability that it does \emph{not} die in a given time interval $\dt$ is given by the following:

\begin{equation*}
P(\sigma_{i}(t+\dt)=I | \sigma_{i}(t) = I) = e^{-\delta \dt},
\end{equation*}

\noindent where we adopt the notation $\sigma_{i}(t) \in \{T,I,D\}$ for the state of cell $i$ at time $t$. Since infected cells can either die or stay infected in a given time step, we conclude that the transition probability is

\begin{equation}\label{eq:inf_to_death_prob}
P(\sigma_{i}(t+\dt)=D | \sigma_{i}(t) = I) = 1 - e^{-\delta \dt}.
\end{equation}

\noindent Considering the transition from uninfected cell to infected cell, $T\rightarrow I$, we apply a similar argument. Here, the rate of infection is dependent on the viral load as well as the infection parameter $\beta$. We assume that cell $i$ can be infected by the quantity of virus in the domain of the cell, $\int_{S_i} v(\mathbf{x},t) d\mathbf{x}$, and that this quantity has negligible change over the small time interval $\dt$. However, by only considering the viral load within cell domain $S_i$, rather than the global domain $\Omega$, we effectively shrink the tissue size by a factor of $\left| S_i \right|/\left| \Omega \right| = 1/N$. Since $\beta$ is a contact parameter and depends on the size of the cell population, we must the rescaling $\beta \rightarrow \beta \nabm$ in order to preserve the dynamics of the original system (for details, see the Supplementary Information). The effective rate of infection is therefore $ \nabm\beta \int_{S_i} v(\mathbf{x},t) d\mathbf{x}$, and using the argument above, we obtain the transition probability for infection:

\begin{equation} \label{eq:target_to_inf_prob}
P(\sigma_{i}(t+\dt)=I | \sigma_{i}(t) = T) = 1 - \exp \left( - \nabm\beta \int_{S_i} v(\mathbf{x},t) d\mathbf{x} \dt \right).
\end{equation}

\noindent Collectively, Eqs \eqref{eq:virus_pde}--\eqref{eq:target_to_inf_prob} govern the agent-based form of the model. We apply periodic boundary conditions on opposite sides of the computational domain to eliminate boundary effects. 

Throughout this work, we will refer to the following parameter set, which is taken from values found in \cite{hernandez_vargas_modelling_covid_in_humans} (with $\beta$ rescaled appropriately for the number of cells in our model). In their paper, Hernandez-Vargas and Velasco-Hernandez fit the TIV model, Eqs \eqref{eq:TIV_T}--\eqref{eq:TIV_V} to data from human COVID-19 patients. Clearly, adapting this model to a spatial context and introducing viral diffusion modulates the behaviour of the system compared to its ODE form, but these parameter values are sufficiently biologically realistic for the purposes of this work.

\begin{table}[h!]
	\centering
	
	\begin{tabular}{p{5cm}|p{1.5cm}|p{6cm}}
		
		\hline
		Description & Symbol & Value and Units  \\
		\hline
		Infection rate& $\beta$ & $1.83 \times 10^{-5} \text{ (copies/ml)}^{-1} \text{ h} ^ {-1}$ \\
		Death rate of infected cells& $\delta$ & $4.33 \times 10^{-2} \text{ h} ^ {-1}$  \\
		Virion production rate& $p$ & $2.23 \times 10^{-1} \text{ cell}^{-1} \text{ h}^{-1}$  \\
		Viral clearance rate& $c$ & $0.1 \text{ h}^{-1}$\\[0.5em]
		\hline
		Number of cells & $\nabm$ & 14,400 (120 $\times$ 120 grid)\\

	\end{tabular}
	\caption{Parameters used in our simulations.}
	\label{tab:tiv_params}
\end{table}

The choice of diffusion coefficient, $D$ has a profound influence on the qualitative outcome of model simulations as well as numerical requirements, as we shall see. However, the question of which diffusion coefficient is most appropriate in a given scenario does not have a straightforward answer \cite{gallagher_et_al_causes_and_consequences}. The kind of spatial data required to calibrate the viral diffusion coefficient is in limited supply, and as such $D$ is usually inferred from other, potentially conflicting sources \cite{gallagher_et_al_causes_and_consequences}. \emph{In vitro} infection assays for most viruses, for example, show the formation of tight viral plaques, with ring-shaped infection fronts, which suggest a fairly small diffusion coefficient \cite{wodarz_et_al_complex_spatial_dynamics_oncolytic, chiem_et_al_covid_reporter_in_vitro, holder_et_al_design_considerations_influenza}. There is also evidence from \emph{in vivo} experiments which suggest highly localised spread of virus. Work by Fukuyama and colleagues, which studied mice infected with multiple differently-coloured fluorescent reporter influenza viruses, showed the formation of single-coloured clusters of infected cells, suggesting virus had spread locally from the first infected cells \cite{fukuyama_et_al_color_flu}. On the other hand, estimations of the diffusion coefficient using the Stokes-Einstein equation, based on viral particle diffusion in mucus, suggests a larger value for $D$ \cite{holder_et_al_design_considerations_influenza, sego_et_al_covid_model}. Moreover, a larger diffusion coefficient would explain long-range viral dispersal through sections of tissue within the host, for example the respiratory tract. These conflicting phenomena make it challenging to fix a value for $D$.

\begin{figure}
		\centering
		
		\includegraphics[width=0.9\linewidth]{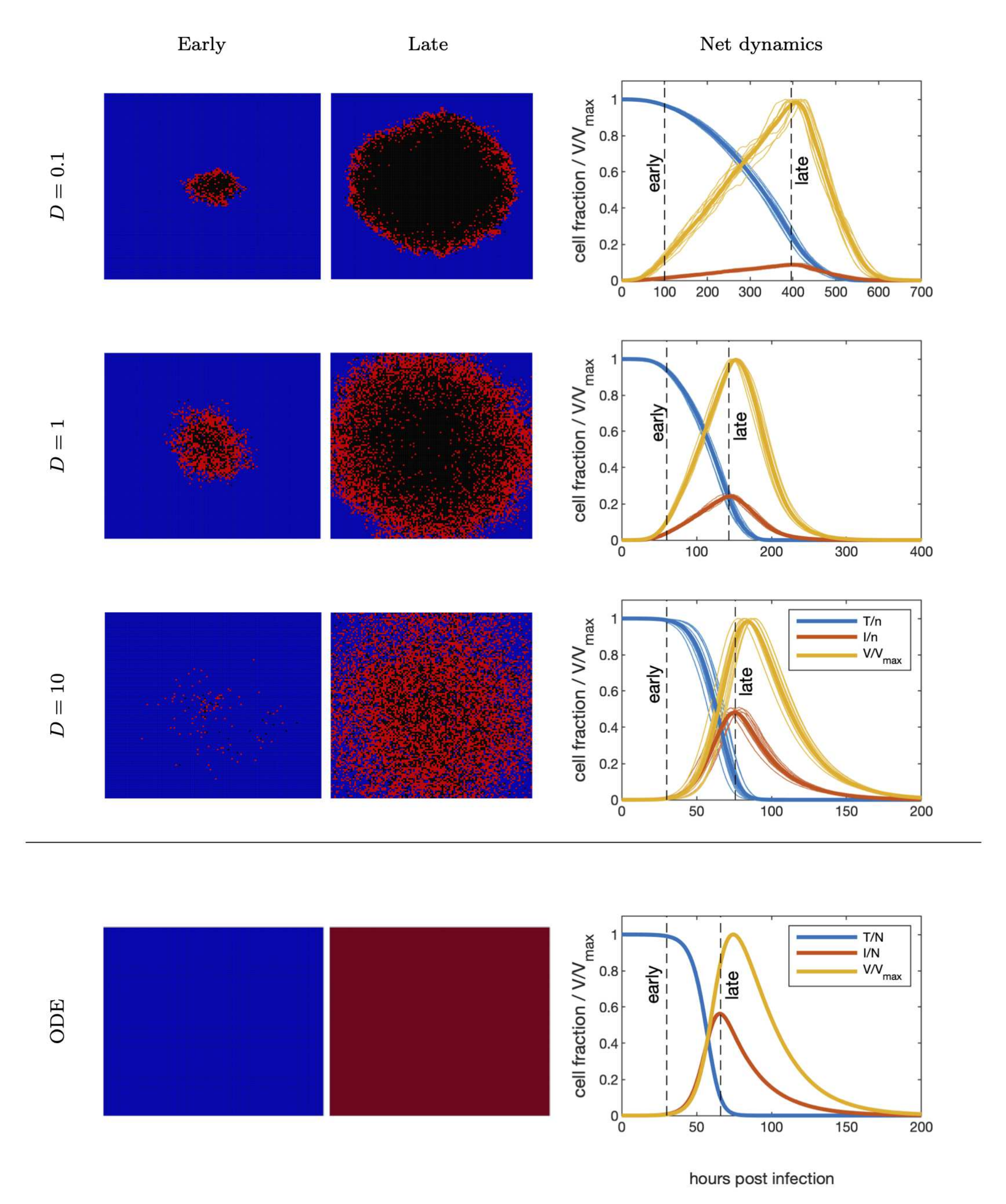}

	\caption{The multicellular TIV model under the parameters in Table \ref{tab:tiv_params} and different diffusion coefficients (units of CD$^2$ h$^{-1}$), compared with the dynamics of the analogous ODE form of the model. For each choice of diffusion coefficient, we initiate the multicellular model with a single infected cell in the centre of the grid. In the left two columns is the state of the cell grid early in the infection and near the peak infected cell fraction. Target cells are shown in blue, infected cells in red, and dead cells in black. For the ODE case, we shade the entire grid according to the proportions of each cell type using the same colour scheme. In the right column we show the net dynamics of the model with the corresponding diffusion coefficient, plotting time series of target and infected cell fractions and total viral load (normalised relative to its maximum). For the multicellular model, we plot time series for ten model simulations and also show the average in bold. Vertical dashed lines represent the ``Early'' and ``Late'' time points for that diffusion coefficient, at which the corresponding cell grid images were taken.}
	\label{fig:diff_cluster_demo_v3}
\end{figure}

In Fig \ref{fig:diff_cluster_demo_v3}, we demonstrate the infection patterns formed from a single source cell under three biologically plausible scales for the diffusion coefficient, and also show the results of the equivalent ODE system, Eqs \eqref{eq:TIV_T}--\eqref{eq:TIV_V}, with the same parameters. Crucially, at small- and medium-scale diffusion ($D=\mathcal{O}(10^{-1} - 10^{0})~\text{CD}^2\text{h}^{-1}$) (where CD is a typical Cell Diameter, here taken to be $10\mu m$ as the size of a typical epithelial cell in the respiratory tract \cite{devalia_et_al_nasal_bronchial_cells}), the infection forms a distinct plaque with infected cells on the periphery and dead cells at the centre. As diffusion increases, this pattern is distorted to become increasingly uniformly distributed. In the case of large diffusion, for example $D=\mathcal{O}(10^1)~\text{CD}^2\text{h}^{-1}$, the ring structure is lost, conditions become well-mixed, and the model behaviour rapidly converges to the ODE solution. We will conduct most of our analysis here using $D=0.1~\text{CD}^2\text{h}^{-1}$ and $D=1~\text{CD}^2\text{h}^{-1}$.

\subsection* {Numerical implementation}

We simulate our model by updating the state of both the cell grid and the virus density every small time step $\dt$. If we set $\mathcal{G}_{\tau}$ to be the state of the cell grid and $\mathbf{v}_{\tau}$ to be the numerical virus density surface at discrete time $\tau$, then with each time step we do the following:

At the start of a given time step at time $\tau$, we begin by checking every cell in the simulation for a state transition and generate the next state of the cell grid $\mathcal{G}_{\tau+\dt}$. We then update the virus density surface using a forward Euler finite difference scheme, based on the state of the grid at the start of the time step, $\mathcal{G}_{\tau}$. We then update both the cell grid and the numerical virus density and increment time by $\dt$. $\dt$ is chosen to satisfy the Courant stability of the PDE based on our range of choices for diffusion and spatial discretisation parameters, i.e the same $\dt$ is used for each simulation.

\begin{figure}[h!]
	\centering
		
\includegraphics[width=0.9\linewidth]{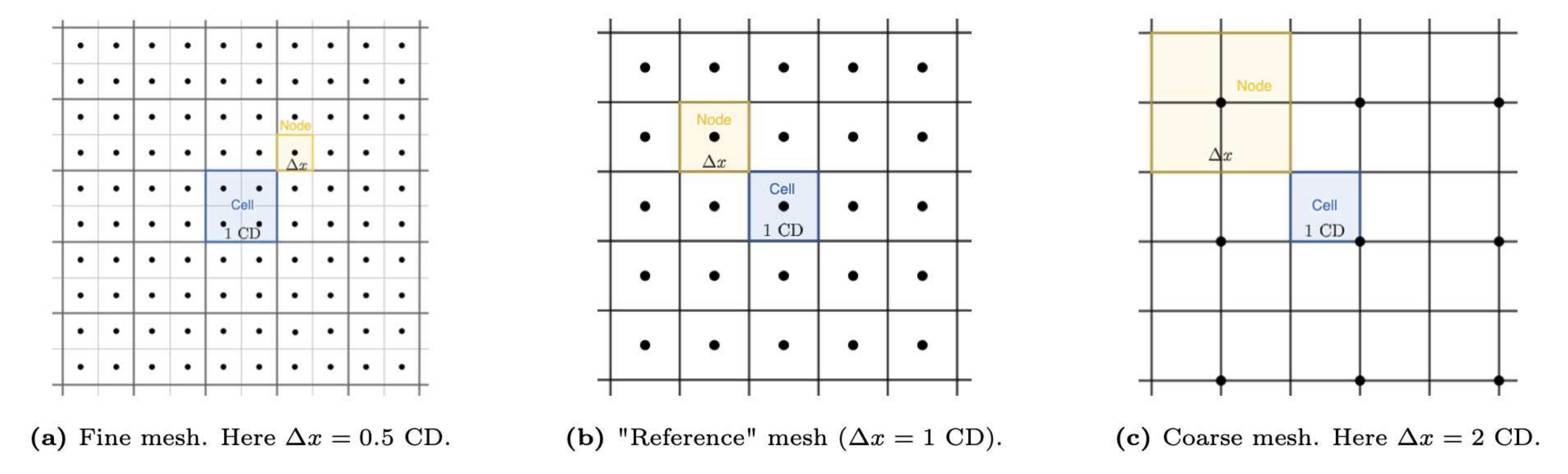}

	\caption{Alignment of the cell grid and virus mesh. Note that in the fine mesh case, each viral node uniquely belongs to one cell each, and in the coarse case, each cell uniquely belongs to one viral node each. In the reference case there is a 1-1 correspondence between cells and nodes.}
	\label{fig:grid_alignment}
\end{figure}

We solve the virus PDE over a finite difference mesh, which is distinct from the cell grid, although both occupy the same spatial domain $\Omega$. We consider cells to be squares of unit length, and hence use cell widths (or \emph{diameters}) as our unit of spatial measurement. We vary the spacing of the nodes of the PDE mesh, $\dx$, to investigate any influence this may have on the results of model simulations. We restrict ourselves to $\dx$ values of the form $k$, or $1/k$, where $k \in \mathbb{Z}$ to avoid any ambiguities with mesh nodes lying on cell edges. That is, we consider mesh spacings that are either a whole number of cell widths apart, or an integer fraction of a cell width apart. Fig \ref{fig:grid_alignment} shows the alignment of the virus PDE mesh relative to the cell grid under these assumptions. For further details of the numerical implementation of our model, see the Supplementary Information.

As this construction suggests, $\dx=1$ is a special case in which the virus surface is discretised with the same spacing as the cell grid (middle plot in Fig \ref{fig:grid_alignment}). Under such a scheme, each cell contains exactly one node. We will refer to this as the ``reference case'' since it is the most conceptually simple scenario where virus is passed not between abstract nodes in space but between actual cells. As such, we will present results of altering $\dx$ compared to this reference.

\section*{Results}
\subsection* {Choice of spatial discretisation affects virus export from infected cells}

The convergence of the diffusion (heat) equation for small values of $\dx$ is a standard result in numerical methods. However, the introduction of an alternate spatial scale --- the cell --- necessitates a more careful discussion of convergence. We sought to understand how the numerical solution of a diffusive process could influence the amount of the diffusing material available at a cell probe, as opposed to a single point. In the context of hybrid viral models, the amount of virus available to cells as it diffuses in the domain is of particular importance, as this will influence the probability of infection.

To investigate how the spread of virus across the tissue might be influenced by the choice of refinement in the spatial discretisation, we tracked how the virus generated by a single infected cell spreads across the cell sheet during its lifetime. To do this, we made the following simplifications to the model: (i) we set a single cell to be infected, and prevented any other cells in the sheet from becoming infected, and (ii) we set viral decay $c=0~\text{h}^{-1}$, such that the evolution of the viral density is determined only by diffusion and its source term. Simulations were initiated with no extracellular virus in the system. Moreover, after the mean infected cell lifetime $1/\delta ~ \text{h}$ (this follows simply from the ODE form of the model, Eqs \eqref{eq:TIV_T} -- \eqref{eq:TIV_V}), we set the source cell to die, such that the source of virus is ``turned off'', and allow more time for the remaining virus to spread in the tissue. 

\begin{figure}
	\centering
			
	\includegraphics[width=0.9\linewidth]{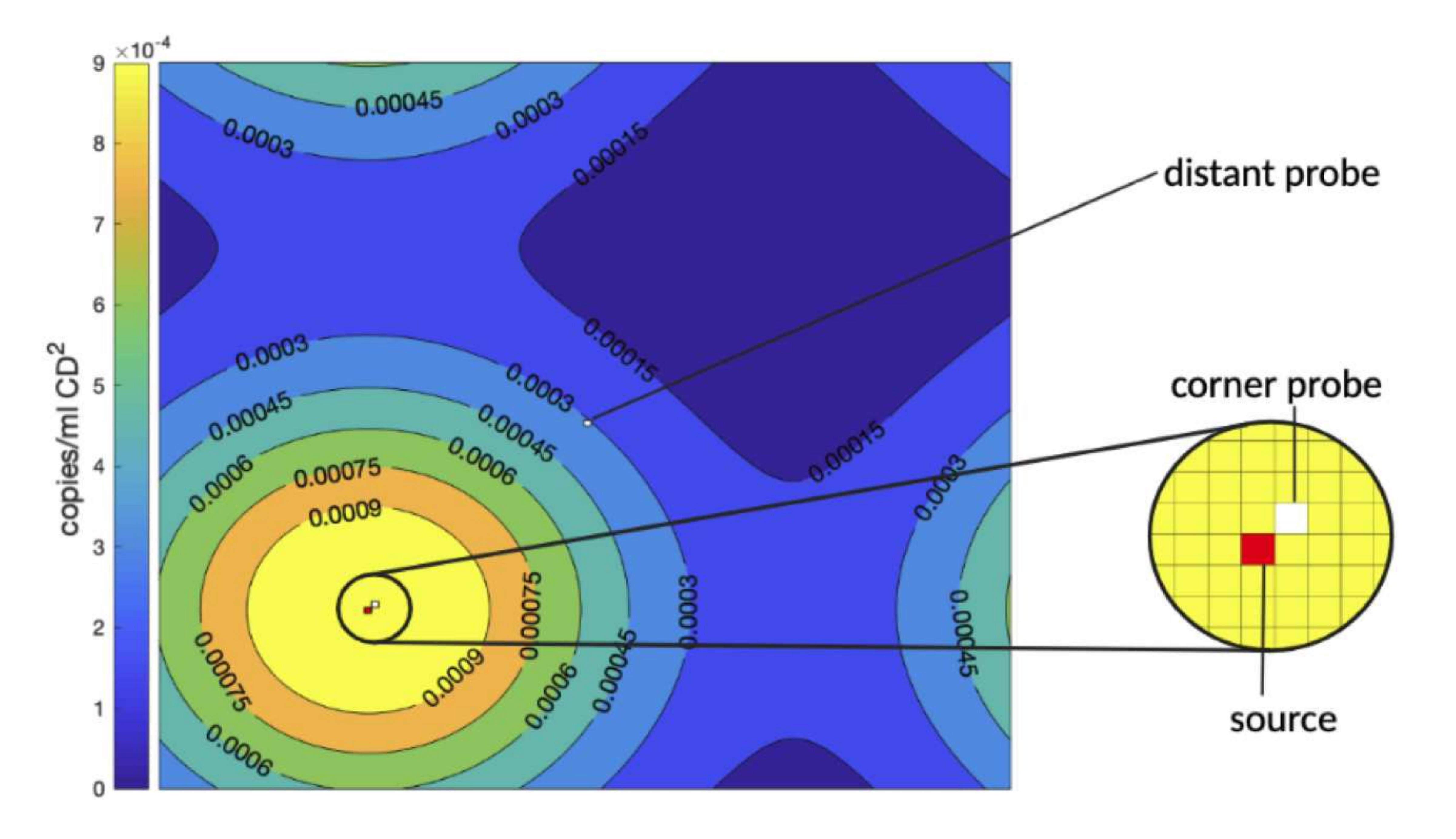}
	
	\caption{Demonstration of probe experiment and the placement of the source and probe cells. We plot the contours of the viral surface after $t=50\text{h}$, using $D=10~\text{CD}^2\text{h}^{-1}$.}
	\label{fig:convergence_cartoon}
\end{figure}

We record the viral load available to two \emph{probe} cells over time under these conditions, as well as at the source cell itself. The two probes are chosen such that one is nearby to the source (they share a corner), and the other is relatively distant (at a diagonal distance of 30 cells). This reflects long- and short-range interactions between cells in the tissue. In Fig \ref{fig:convergence_cartoon} we show a diagram of the virus density surface after $50\text{h}$ and indicate the location of the probes (note: for ease of presentation, Fig \ref{fig:convergence_cartoon} was generated using $D=10~\text{CD}^2\text{h}^{-1}$; in the following analysis we will take $D=0.1~\text{CD}^2\text{h}^{-1}$). We repeated this experiment for a range of $\dx$ values. In cases where the PDE mesh is coarser than the cell grid ($\dx >1$), the position of the source cell relative to the configuration of the cell grid biases the amount of virus available at the probes. For example, if $\dx=2$ as in Fig \ref{fig:grid_alignment}, the diagonal neighbour (the near probe) of a given cell may correspond to the same viral node, but may also correspond to a neighbouring viral node. To account for this artefact in these cases, we chose the source cell at random and identified probe cells relative to their position, and reported the average behaviour over 200 such simulations.

\begin{figure}
	\centering
			
	\includegraphics[width=0.9\linewidth]{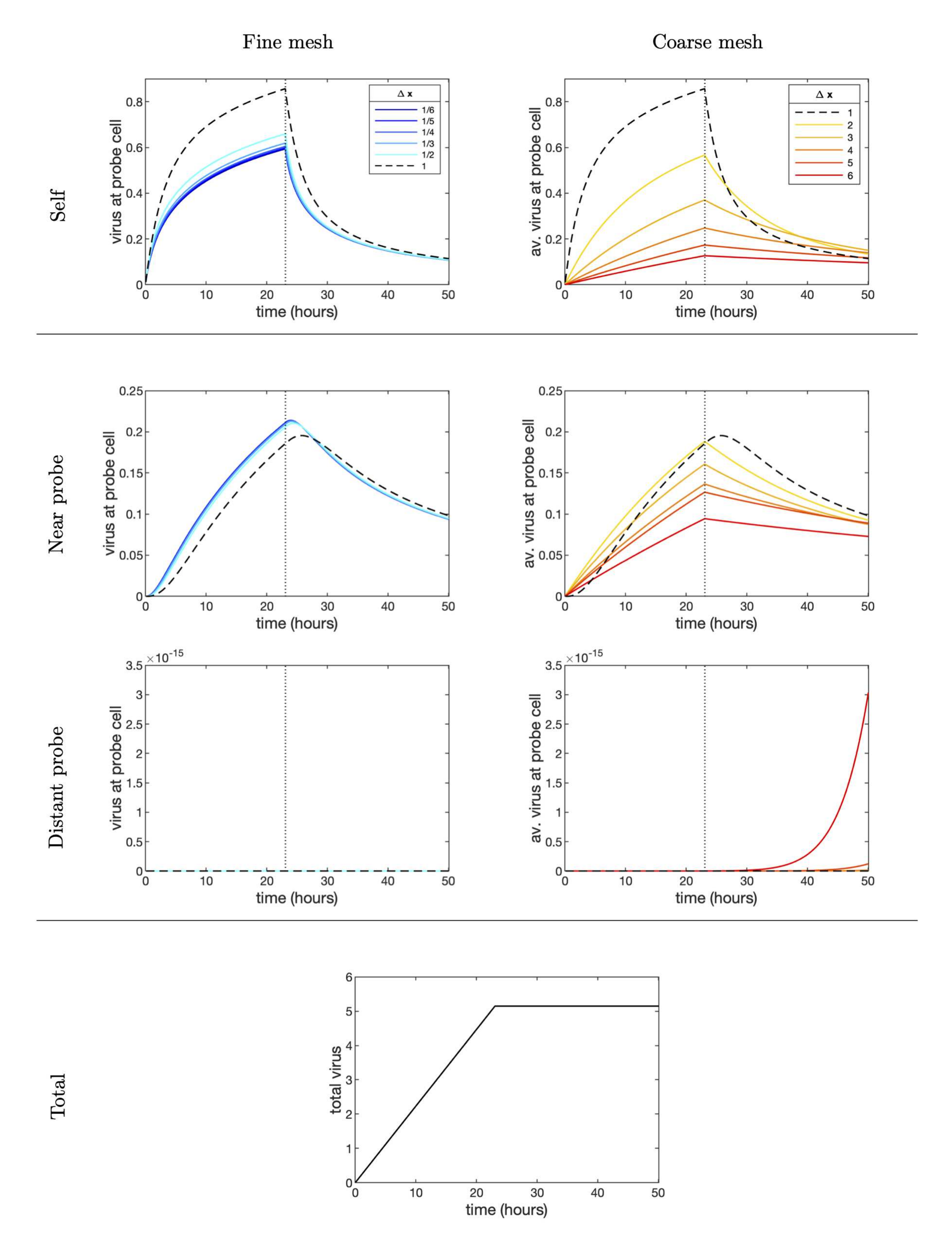}
	
	\caption{Effect of choice of spatial discretisation on virus availability at two probe cells both near to and distant from a single infected source cell, as well as at the source cell itself, using $D=0.1~CD^2h^{-1}$. We also plot the total viral load in the system over time, which is identical for any choice of $\dx$. In the case of coarse mesh refinement, we choose the probe cell at random to account for the position of the source and probe cells relative to the viral nodes, and take average behaviour across 200 simulations.}
	\label{fig:convergence_fig_main}
\end{figure}

Fig \ref{fig:convergence_fig_main} shows the outcome of this experiment, using a diffusion coefficient of $D=0.1~\text{CD}^2\text{h}^{-1}$. In each plot, we chart the viral load at the indicated cell over time under different choices of $\dx$ values. In the left column of Fig \ref{fig:convergence_fig_main}, we show the behaviour of the system for increasing refinement of the PDE discretisation ($\Delta x \rightarrow 0$, i.e. convergence of the model). From the plots of viral load at the near probe and the source cell, we notice that, whilst the dynamics converge as $\dx \rightarrow 0$, there is a significant deviation between the limit of mesh refinement and the reference case of $\dx=1$, at least under our choice of a small diffusion coefficient. Our results suggest that increasing mesh refinement leads to less virus remaining at the source cell, and increases the amount of virus found at nearby cells. Note that this effect is observed even though the amount of virus produced in each case is identical, regardless of the value of $\dx$ (see Figure~\ref{fig:convergence_fig_main} Total). At the distant probe, the viral load remained negligibly low throughout the entire simulation for any value of $\dx\le1$.

In the case of coarse mesh refinement (right column of Fig \ref{fig:convergence_fig_main}), model simulations rapidly diverge from the reference case ($\dx=1$) as $\dx$ increases. In particular, the viral load time series at the near probe is qualitatively different to the reference case, even for $\dx=2$. At the distant probe, it can be seen that in the very coarse case, proportionally much more virus reaches the cell towards the end of the simulation. Unlike with the near probe, this divergence arises long after the source cell has died, indicating that this viral load occurs from the virus already in the system equilibrating across the sheet. This process is evidently accelerated when $\dx$ is very large, however, is unlikely to influence the simulation in any meaningful way, since the actual \emph{amount} of virus at this distance from the source cell is still extremely low (on the order of $10^{-15}$ copies/ml compared to $\mathcal{O}(10^{-1})$ copies/ml at the near probe).

\begin{figure}[t]
	
	\centering

\includegraphics[width=0.9\linewidth]{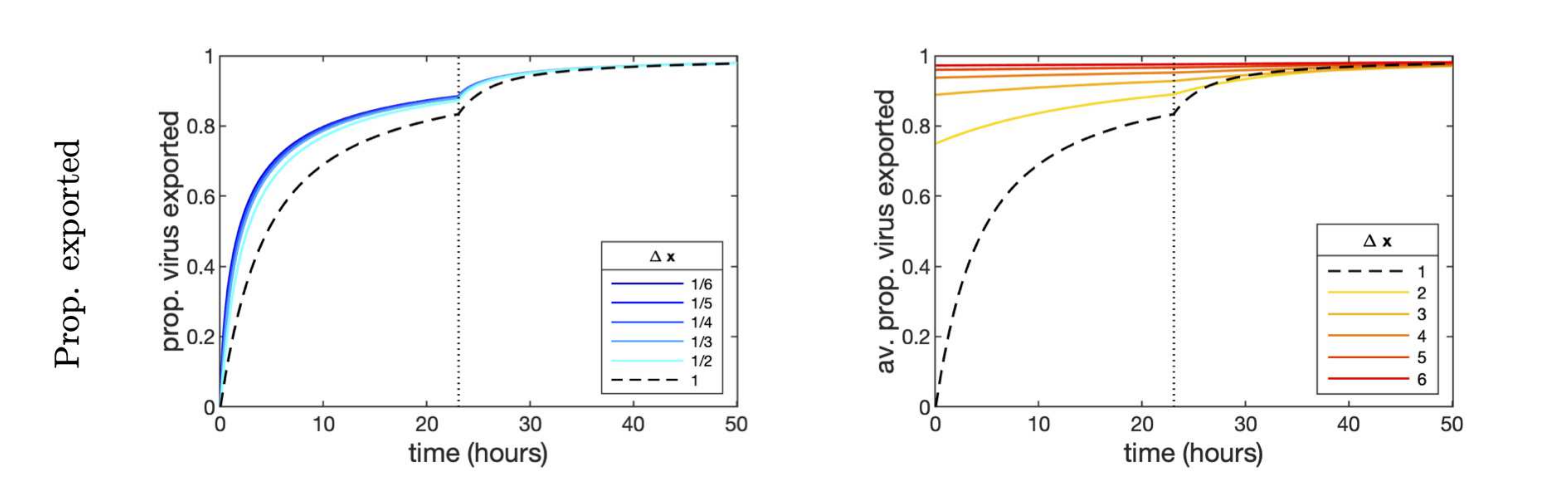}
	
	\caption{Proportion of ``productive'' virus in system (proportion of virus external to the source cell). Here we use $D=0.1~CD^2h^{-1}$.}
	\label{fig:prop_exported}
\end{figure}

As with the fine mesh case, coarse mesh refinement leaves significantly less virus at the source cell, however, here, nearby cells also tend to have access to less virus. Fig \ref{fig:prop_exported} provides insight into the mechanics of this viral export process, and how it is affected by the choice of $\dx$. Here, we plot the proportion of the total virus in the system which is \emph{not} found at the source cell. That is to say, we plot the proportion of virus which has been exported from the source cell. This may be considered the proportion of \emph{productive} virus in the system at a given time: given that the source cell is already infected, any virus which remains in its vicinity cannot produce any new infections. Fig \ref{fig:prop_exported} shows that the proportion of productive virus is greater than the reference case for both fine and coarse discretisation schemes. This suggests that virus is more rapidly and effectively transported from the source cell in either case, despite none of the actual model parameters being changed.

As Fig \ref{fig:prop_exported} suggests, there are separate mechanisms which explain this enhanced transport in both the fine and coarse mesh cases. Where the PDE mesh resolution is coarser than the cell grid, any virus generated by an infected cell is immediately accessible to all other cells which share the same viral node. As such, only a small proportion of the virus generated by the source cell remains at the source, and the rest instantly becomes productive. This phenomenon explains the substantially and instantaneously higher proportion of exported virus for coarser mesh resolutions in Fig \ref{fig:prop_exported}. Moreover, as the mesh becomes increasingly sparse, more cells share in the instantaneous spread of produced virus, and the resulting viral distribution becomes broader and flatter. This explains the flattening viral load curve at the near probe and the increasing viral availability at the distant probe in Fig \ref{fig:convergence_fig_main}.

The acceleration of viral transport in the fine mesh case is more subtle and does not act instantaneously. Here viral export is enhanced because, relative to the case where $\dx=1$, increasing mesh refinement reduces the distance virions must travel from a cell boundary to reach adjacent cells. If cells each contain only a single viral node, virions need to travel an entire cell width to access neighbouring cells, however, if cells contain many nodes, virions need only travel a fraction the width of a cell to access the viral nodes attached to neighbours, thereby reducing the time before cells begin to experience a concentration of virus. Clearly, this ``spill-over'' effect is enhanced when the diffusion coefficient $D$ is small enough that viral spread across sub-cellular length scales occurs over a non-trivial time interval. In the Supplementary Information we show that this effect is attenuated as $D$ increases. We remark here, however, that should a modelling application demand a diffusion coefficient on the order of what we have shown in Fig \ref{fig:prop_exported} (such that very fine mesh resolution ($\dx < 1$) is necessary to ensure convergence), the assumptions of the multi-scale model begin to break down. For example, at such a fine level of spatial detail, the packing of cells (which is closer to hexagonal than rectangular \cite{blahut_et_al_hepatitis_c_two_modes_of_spread, graw_et_al_hcv_cell_to_cell_stochastic}) and heterogeneity in receptor availability start to become important, and even the assumption that virus is well-approximated by a density surface begins to come under strain.

\subsection* {Choice of spatial discretisation influences the computed reproduction number}

Having shown that changing $\dx$ modulates the spread of virus across the tissue, we sought to demonstrate how this affects the ``biology'' of the model. In the Methods Section we discussed the notion of the basic reproduction number $\rzero$, which characterises the TIV system of ODEs. $\rzero$ measures the number of infections caused by a single infected cell over its lifetime, and is therefore a useful measure of viral infectivity, given the parameters of the model \cite{perelson_hiv_review, beauchemin_and_handel_influenza_review}. Such a notion is much more challenging to define in a spatial context, since the availability of target cells is constrained by the spatial arrangement of the tissue. This is further complicated by our use of a discrete cell grid. To circumvent much of this complexity, we here introduce an \emph{empirical} $\rzero$ . This quantity, for which we write $\rzero^*$, is simply defined as the numerically computed number of infections caused by an infected cell over its lifetime under the model. 

To investigate how the choice of spatial discretisation influences $\rzero^*$, we ran simulations of our model using different choices of $\dx$. In each case, we randomly chose 1\% of the cells to begin the simulation infected, and at $t=1/\delta$ all infected cells were killed, having reached their average life span, as in the previous experiment. We allow sufficient extra time for any virus in the system at this point to cause further infections as it spreads and decays, until the probability of further infections becomes negligible. Infection events are implemented here as follows: whenever our model reports a cell becoming infected (according to the state transition probability rules), that cell is flagged to prevent becoming infected again, but does \emph{not} go on to secrete virus. In this way, all infections in the system can be exclusively attributed to the initially infected cells. We calculate $\rzero^*$ as the mean number of infection events per starting infected cell. We repeated the experiment for both $D=0.1~\text{CD}^2\text{h}^{-1}$ and $D=1~\text{CD}^2\text{h}^{-1}$.

We use multiple initially infected cells instead of a single cell for our calculations because it substantially reduces the random variation between simulations. The proportion of the cell sheet which is initially infected has been termed the multiplicity of infection or MOI (in the sense used by, e.g., Whitman \emph{at al.} \cite{whitman_et_al_spatiotemporal_host_virus_influeza}, which differs slightly from how the term is used in the biological literature). Clearly our formulation of $\rzero^*$ depends on the choice of MOI (which we explore in the Supplementary Information), however, it is sufficient to show how the viral infectivity of the model is affected by change in the spatial discretisation.

\begin{figure}[h!]
	
	\centering

	\includegraphics[width=0.9\linewidth]{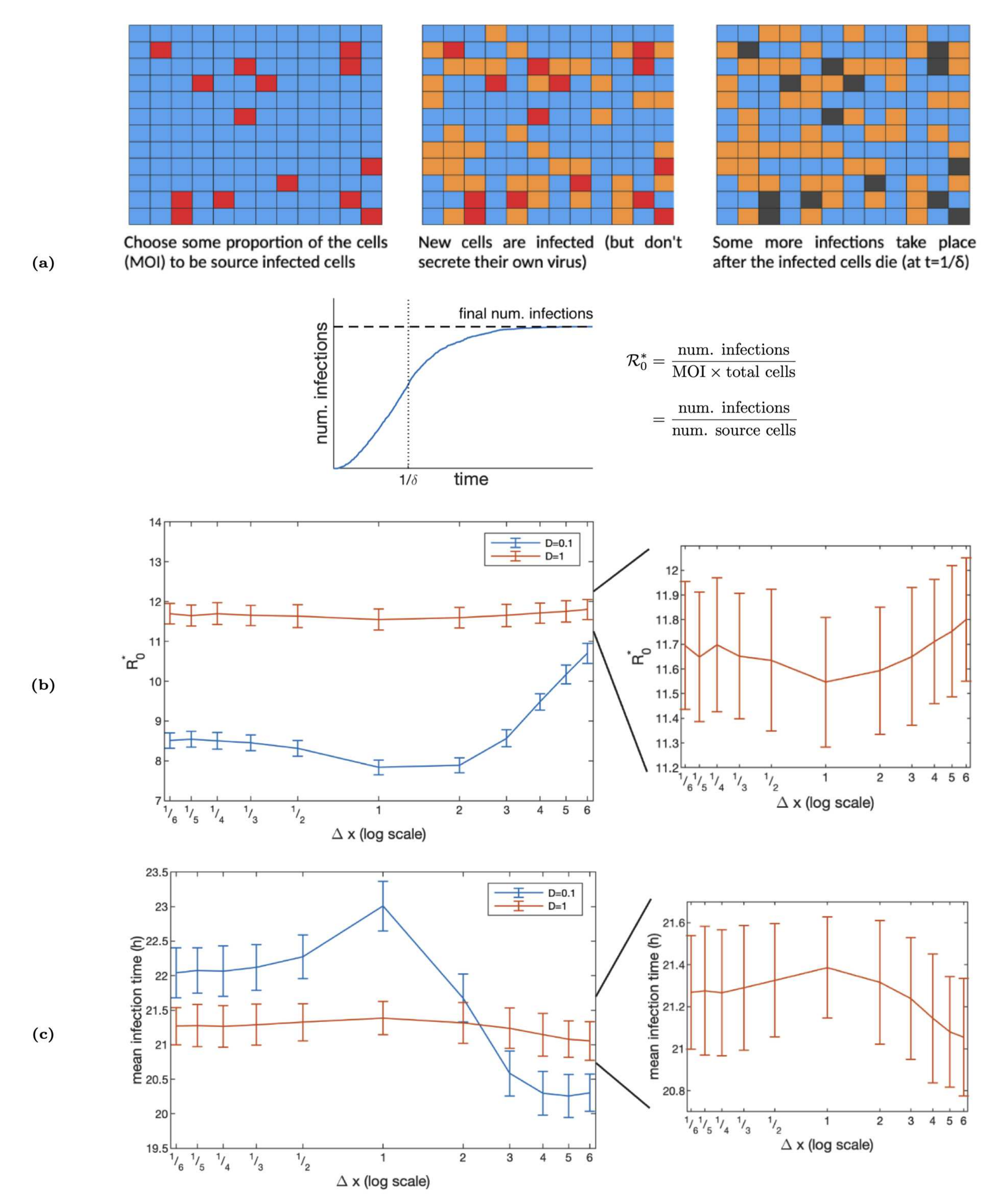}
	
	\caption{Spatial discretisation affects the (empirical) basic reproduction number of the model, and causes infections to occur earlier on average. (a) Derivation of the ``empirical'' basic reproduction number, $\rzero^*$. This quantity measures the average number of infections caused by an infected cell over its lifetime, including infection by virus left in the system after the death of the infected cell. (b) $\rzero^*$ for different choices of mesh refinement ($\dx$) and diffusion coefficient $D$. We use a 120$\times$120 grid of cells and randomly select 1\% of cells as initially infected (MOI=0.01). We report the mean results of 200 simulations. Error bars indicate one standard deviation. (c) Mean infection times for the simulations in (b).}
	\label{fig:rzero_full_fig}
\end{figure}

In Fig \ref{fig:rzero_full_fig}(b), we plot $\rzero^*$ values for different choices of $\dx$ alongside error bars indicating one standard deviation. Fig \ref{fig:rzero_full_fig}(b) shows that this ``basic reproduction number'' is increased both when $\dx$ is smaller and when it is larger than the reference case at $\dx=1$. This extends our findings from the previous result, by demonstrating that the accelerated transport of virus in both very fine and very coarse mesh schemes results in an increased number of infections. Consequently, the model predicts a higher viral invasive potential, \emph{despite none of the actual model parameters changing}.

Again, this effect is stronger when diffusion is smaller. In the case where $D=0.1~\text{CD}^2\text{h}^{-1}$, the variation in $\rzero^*$ is substantial and vastly exceeds the noise present in the system. When $D=1~\text{CD}^2\text{h}^{-1}$, there is still an appreciable change in the basic reproduction number for varying $\dx$, however, this variation is proportionally less significant relative to the randomness of the system.

In Fig \ref{fig:rzero_full_fig}(c), we also plot the mean of the infection time distribution for each choice of $\dx$.  Fig \ref{fig:rzero_full_fig}(c) shows that the mean infection time is reduced compared to the reference case in both fine and coarse discretisation schemes. This result follows from the finding in Fig \ref{fig:convergence_fig_main} that viral export is accelerated in these settings, and complements our $\rzero^*$ result by demonstrating that small and large $\dx$ values lead to not only more infections, but also, on average, earlier infections. 

However, this increase in the number of infections (reproduction number) cannot fully be explained by simply adding infections that occur earlier in the simulation. For example, comparing the $\dx=1$ reference case with the $\dx=2$ case, note that whilst the average infection time is significantly reduced for the coarser mesh, this barely results in any change in $\rzero^*$. This suggests that whilst early infections are more likely under this scheme, there must also be a \emph{reduction} in the likelihood of infections at later times. These results point to a complex relationship between the distribution of infection times and the spatial discretisation. 

We explore this relationship in the Supplementary Information. Our analysis suggests that in coarse mesh schemes, the early acceleration in infection is driven by infections of cells which share a viral node with a cell which produces virus. This eliminates viral transport time and permits rapid infections early in the simulation. However, compared to large $\dx$ cases, where there are many target cells which share viral nodes with initially infected cells, when $\dx$ is only slightly larger than 1, this pool is much smaller. Once availability of these cells becomes constrained, the infection takes longer to reach cells further away than in the reference case, and the infection slows down. This could explain the plateau in $\rzero^*$ for $\dx=1,2,3$, while the mean time of infection notably dips (see Supplementary Information for details).

\subsection* {Choice of spatial discretisation influences the time scale of the viral dynamics}

So far, we have studied simplified versions of the model in order to pinpoint processes which are influenced by the choice of mesh refinement. We now shift our attention to the full model to examine the effect of spatial discretisation on full, biologically relevant simulations. As a metric by which to compare different simulations, we chose the time of the peak viral load. This quantity is relatively consistent for a given set of parameters (compared to the time to viral extinction, e.g., which is subject to considerable noise) and provides a notion of the aggressiveness of the infection as described by the model. For a range of choices of $\dx$, we initiated simulations with MOI=0.01 and took the average viral peak time of 32 simulations. As before, we repeated the experiment for both $D=0.1~\text{CD}^2\text{h}^{-1}$ and $D=1~\text{CD}^2\text{h}^{-1}$.

\begin{figure}[h!]
	\centering

	\includegraphics[width=0.9\linewidth]{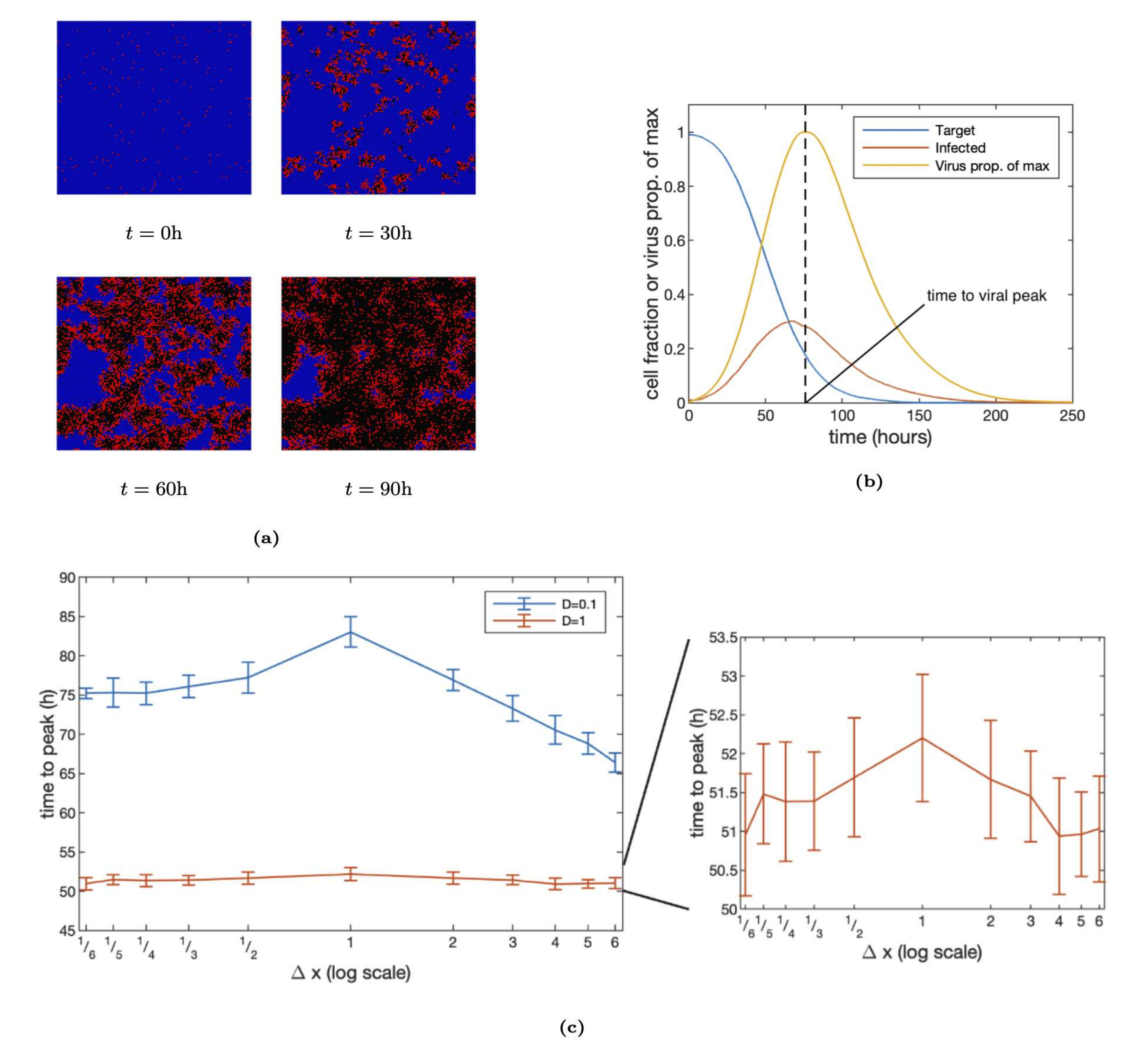}

	\caption{Time to peak viral load is affected by the choice of spatial discretisation. (a) Cell grid at selected time points in a typical simulation at MOI=0.01 ($D=0.1~\text{CD}^2\text{h}^{-1}$). Target cells are blue, infected cells are red, and dead cells are black. (b) Net dynamics of the simulation shown in (a). We plot the proportion of cells which are target cells (blue) or infected (red). The total viral load in the system is shown relative to its maximum in yellow. The dashed line represents the time of peak viral load. (c) Time of peak viral load as influenced by choice of $\dx$ for different choices of diffusion coefficient. Here, in each case, we initiate the simulation with MOI=0.01 and perform 32 simulations for each choice of $\dx$ to account for randomness in the model. Error bars reflect one standard deviation either side of the mean.}
	\label{fig:peak_times}
\end{figure}

Fig \ref{fig:peak_times} shows a plot of the average peak times against values of $\dx$. Error bars indicate one standard deviation. This plot shows that, provided the diffusion coefficient is sufficiently small, the choice of spatial discretisation can have a substantial impact on the time to peak. Our plot of the mean peak times show that in the small diffusion case, the $\dx=1/6$ and $\dx=6$ cases vary by approximately 9 and 19\% respectively compared to the reference case, $\dx=1$.

As in the previous result, Fig \ref{fig:peak_times} shows that for time to peak, the reference case is an extremum. Both increasing and decreasing the resolution of the mesh results in an acceleration of the infection. This follows from our result that the basic reproduction number of the system is effectively increased in these cases, which in turn is a consequence of our finding that in fine and coarse mesh schemes, viral export from infected cells is enhanced.

Once again, this effect is attenuated when the diffusion coefficient increases, especially relative to the randomness of the system. However, as the inset of Fig \ref{fig:peak_times} shows, even when the diffusion coefficient is larger ($D=1~\text{CD}^2\text{h}^{-1}$), a change in $\dx$ still causes an appreciable and consistent variation in the time to peak according to the same trend as in the small diffusion case.


\section*{Discussion}
Spatially-realised models of viral infections within the host represent an opportunity to account for biological detail not possible in ODE models, however, representing the two spatial scales of cells and virus (or other small particles) represents a conceptual challenge \cite{perelson_and_ke_covid_review, gallagher_et_al_causes_and_consequences}. One popular and ``natural'' approach to account for these two spatial scales is by coupling a discrete grid of cells to a virus density surface \cite{gallagher_et_al_causes_and_consequences, sego_et_al_cellularisation}. Simulating such a system, however, requires the discretisation of the viral density. Here, using a simple but robust model derived from standard ODE models of viral dynamics, we have provided a detailed analysis of the effects of this discretisation on model outcomes, subject to various modelling conditions. 

In particular, we examined how the behaviour of the model was changed relative to the behaviour of  the reference case, where the viral surface is discretised according to the same spacing as the cell grid (that is, where there is one discrete viral node per cell). We showed that, relative to the reference case, both very fine and very coarse mesh refinements both act to accelerate the rate of infections in the model. Note that in the case of refining the mesh this acceleration decreased for increasing refinement leading to convergence whereas for coarsening meshes this acceleration continued increasing with increased coarsening. In Fig \ref{fig:prop_exported}, we showed that in both of these cases, the rate at which virus leaves the cell from which it is secreted is increased, despite the actual amount of released virus being identical. There are two distinct mechanisms for this behaviour. In the coarse mesh case, the increased rate of viral export is simply explained by the presence of multiple cells which share each viral node (a \emph{plate} of cells). Each cell within a plate has instantaneous access to virus released from any other cell on the plate, which accelerates transport between local cells and speeds up infection. In the fine mesh case, we hypothesise that the increased rate of viral export is due to the reduction in distance that virus needs to travel to gain access to neighbouring cells. Compared to the reference case --- where virus must travel from a viral node at the centre of the secreting cell to another node at the centre of a neighbouring cell --- when the mesh is very refined, virus need only reach viral nodes just beyond the cell boundary before that cell experiences viral load. 

These two mechanisms impact the ``biology'' of the model in a complex and non-trivial manner. We introduced the notion of an empirical $\rzero^*$ --- analogous to the basic reproduction number $\rzero$ from ODE theory on viral dynamics --- to report the number of infections generated by a single infected cell over its lifetime, according to our model and a given set of conditions. We showed that, predictably, $\rzero^*$ was increased relative to the reference case when the mesh was coarsened or refined, however, our analyses revealed that there was minimal increase in the number of infections for $\dx$ values only slightly larger than the reference. This was despite the fact that the average infection time in these simulations was distinctly earlier, and was likely due to constraints on the availability of cells which share a viral node with an initially infected cell.

We also showed the impact that these numerical artefacts have on realistic simulations of our model, using the time to peak viral load as a metric. We found that, once again, compared to the reference case, both fine and coarse meshes led to an earlier time to peak, and therefore suggested a more aggressive infection. Interestingly, here, unlike with $\rzero^*$, there is a clear acceleration in the time to peak even for slightly coarse meshes. This suggests that even though the actual number of infections caused by each infected cell (reproduction number) is much the same, the earlier mean infection time means that new generations of infected cells arise more quickly, and the course of the infection is faster nonetheless. This shows that the relationship between $\dx$ and model outcomes is more complex than merely an increase in infection rates. However, regardless of the mechanism responsible, this change in the time to peak is an important finding of our analysis, since we have shown that if the diffusion coefficient $D$ is small enough, the choice of spatial discretisation can interfere with biologically important outputs of the model, and could have consequences for model predictions and parameter estimations. 

The importance of the viral diffusion coefficient, $D$, has been a recurring theme of our results. In the analysis that we have presented here, we have used two scales of diffusion (the exact values of $D$ are of less importance than their order of magnitude and qualitative influence on the model) and shown repeatedly that in the small diffusion case, the model is much more sensitive to $\dx$. If the diffusion coefficient is small enough, care must be taken with the choice of spatial discretisation to ensure that numerical solutions converge: our results show that the reference case where viral nodes have the same spacing as cells may not be sufficiently close to the $\dx \rightarrow 0$ limit. Moreover, in such a scenario, the entire multiscale framework begins to come under strain. We proposed above that the mechanism driving this divergence between the reference case and fine mesh cases was due to a ``spill-over'' effect, where, if $\dx$ is very small, viral export from an infected cell to its neighbours can be accelerated by reaching nodes only just beyond the cell boundary, as opposed to the the single node at its centre in the reference case. Clearly this effect is dependent on the diffusion coefficient being small enough that virus transport over a fraction the width of a cell takes place over a non-trivial time interval. However, should a multiscale model require this level of spatial refinement to converge, this would require a great deal of trust in the spatial arrangement of the cell grid in the model, for example, or in the distribution of viral receptors. Essentially, having very small diffusion under this model framework requires a high degree of detail in the spatial representation of the tissue, to an extent which may exceed the assumptions of a multiscale model.

The values for the diffusion coefficient which we have presented here, $D=0.1~\text{CD}^2\text{h}^{-1}$ in particular, are at the lower end of the range of biologically realistic values (given the conditions of our model, including tissue size, model components etc.) \cite{gallagher_et_al_causes_and_consequences, holder_et_al_design_considerations_influenza, sego_et_al_cellularisation}. As such, the magnitude of the numerical artefacts which we have shown here are somewhat of a worst case scenario for a realistic model. However, we have also shown that even for diffusion one order of magnitude higher, $D=1~\text{CD}^2\text{h}^{-1}$, the choice of spatial discretisation still has a noticeable effect on various outcomes and behaviours of the model, and should be taken into account when considering quantitatively sensitive analyses such as parameter estimation. For larger values of diffusion, for example $D=10~\text{CD}^2\text{h}^{-1}$ (see the supplementary figures), it is true that the choice of spatial discretisation begins to have a negligible effect on the dynamics, at least for the values of $\dx$ which we consider here. However, in exchange for this simplicity, most of the spatial ``information'' of the model is lost. Fig \ref{fig:diff_cluster_demo_v3} shows that for this scale of diffusion, the familiar ring-structured infection pattern is gone, and in fact there is little spatial structure to the infection at all. The net dynamics of the model with $D=10~\text{CD}^2\text{h}^{-1}$ are almost identical to that of the analogous ODE version of the model, which suggests there is little value in carrying around the added complexity of the spatial model.

It is this dependence on the diffusion coefficient being small which limits the scope of our findings fairly specifically to viral dynamics. Other modelling contexts where cell dynamics are coupled with diffusing species, of which tumour spheroid models are a good example (e.g., \cite{ghaffarizadeh_et_al_physicell, cytowski_and_szymanska_timothy_paper}), typically consist of diffusing particles much smaller than virions such as oxygen or signalling molecules which diffuse at much larger rates \cite{sego_et_al_covid_model}. However, unlike the model we have discussed here, even with large diffusion coefficients such models are still spatially structured due to the presence of absorbance or uptake terms \cite{cytowski_and_szymanska_timothy_paper, sego_et_al_covid_model}. Such terms are different to the model framework we have discussed here, and as such our results would need to be reformulated to extend to such systems.

In this work we have only considered one size of the tissue, comprising 14,400 cells. There are several reasons for this choice. For one, this domain size is typical for models of the type we discuss here \cite{sego_et_al_cellularisation, sego_et_al_covid_model, getz_rapid_community_driven_covid_model, blahut_et_al_hepatitis_c_two_modes_of_spread}, and moreover, for the kinds of parameter sweeps and simulation repetitions necessary for this work, having a much larger domain would incur very large computational costs. Furthermore, for a tissue, say, one order of magnitude larger, the multiscale framework (coupled PDE -- discrete cell grid) starts to become unwieldy, if not redundant. Aside from being difficult to visualise and interpret, the huge scale difference between individual cells and the tissue as a whole lessens the significance of the dichotomy between discrete cells and continuous virus density when viewed at a macro scale.	For a tissue of this size, it may be worth considering approximating the tissue by a continuum surface, or bundling cells into discrete patches. Nonetheless, having a substantially larger cell grid would have interesting consequences for our analysis. For example, this would permit larger values of the viral diffusion coefficient with non-trivial spatial effects, since, even though the infection front would not be ring-shaped, it would take some time to spread across the entire sheet. The role that spatial dicretisation of the viral surface would play under such a scheme is not clear \emph{a priori}.

This work has so far considered the Forward Time Centred Space (FTCS) finite difference method for simulating our model, chosen for its simplicity. While it would be infeasible to perform the analysis we have discussed here for all solvers available, the results presented would follow for other numerical schemes. This is because the mechanisms which we have identified as the likely causes of viral acceleration compared to the reference case (namely, the spill-over effect we described for fine discretisation and the instant transport effect between cells sharing a viral node in the coarse case) are relatively method-agnostic. They depend only on the resolution of spatial detail, and not on the actual representation of spatial structure. 

Recent work in the viral dynamics literature has identified the need for a greater understanding of model design in spatial models of infection. Holder and colleagues conducted a detailed study of, among other features, the value of the viral diffusion coefficient in multicellular models of viral infections \cite{holder_et_al_design_considerations_influenza}. Their results characterise spreading infection plaques in terms of model parameters - including the viral diffusion coefficient - and provide some first steps in techniques to estimating these parameters from experimental data. Gallagher and colleagues provided a detailed review and motivation for spatially-resolved models of viral dynamics and used a first-principles approach to advise model selection and design \cite{gallagher_et_al_causes_and_consequences}. They considered a broad range of effects and mechanisms driving spatial spread of infection and provided detailed analysis based on biological data, however stopped short of discussing specific model implementation. More recently, work by Sego and collaborators has shed light on the connection between multicellular viral dynamics models and their ODE counterparts \cite{sego_et_al_covid_model, sego_et_al_cellularisation}. Their work has provided further insight into the role of viral diffusion in these models and developed theory on the changes in model dynamics with the inclusion of spatial structure in viral dynamics models. However, none of these works consider the exact role of spatial discretisation in model outcomes, which, as has been suggested in the literature on cardiac electrophysiology modelling, can have important consequences for model outcomes, from a qualitative perspective and not just a technical one \cite{niederer_et_al_cardiac_tissue_review, fenton_et_al_spiral_wave_breakup}. The results we have presented here demonstrate, quantify and analyse this role in the context of multicellular viral dynamics modelling. We hope that our work continues to improve understanding of the design and study of spatial models of viral dynamics, and can inform the development of powerful modelling tools in future.

In this work, we have explored the role that spatial discretisation plays in the dynamics of multi-scale models of viral infections. We have demonstrated the influence that numerics can have on biologically-relevant outputs of these models and explored the mechanisms which drive these artefacts. Our work will hopefully provide guidance on the construction of spatial models of viral infections and identify structural concerns which may arise. In particular, we showed here that for small enough diffusion, the behaviour of multiscale models may be modulated by their spatial implementation, and, in the worst case scenario, require such a degree of spatial resolution to converge that the model structure may no longer be appropriate. Future work will continue in this trajectory and examine how spatial structure and spatial spread of viral infections are represented in mathematical models.


\section*{Supporting information}

\paragraph*{S1 File.}
\label{S1_File}
{\bf Supplementary Information} Further analysis and figures.

\section*{Acknowledgments}
TW is supported by an Australian Government Research Training Program (RTP) scholarship. JM's research is supported by the ARC (DP170103076, DP210101920)
and ACREME.

%
%
%

\end{document}


\begin{frontmatter}
		\title{Choice of Spatial Discretisation Influences the Progression of Viral Infection within Multicellular Tissues\\[1em] \textbf{Supplementary Information}}
		
		\author{Thomas Williams, James McCaw, James Osborne}
	\end{frontmatter}

	\section{Further details of numerical implementation}

	The choice of spatial discretisation for the virus PDE, $\dx$, has implications for the way in which integrals and virus production is accounted for. In the \textbf{coarse mesh} case ($\dx>1$), it is not possible for cells to secrete virus only in their spatial domain, so secreted virus is instead added to the nearest node, and its value is scaled to account for the larger volume of virus this would generate. That is, if cell $i$ secretes $p\dt$ virus in a given time step, and assuming that virus node $(p,q)$ is nearest to cell $i$, we write:
	
	\begin{equation}
	\mathbf{v}_{p,q}^{t+1} = \mathbf{v}_{p,q}^{t} + \frac{p\dt}{\dx^2}.
	\end{equation}
	
	\noindent The reverse process in this case is relatively simple. The virus available to cell $i$ under this scheme is simply
	
	\begin{equation}
	\int_{S_i} v(\mathbf{x},t) d\mathbf{x}  \approx  \mathbf{v}_{p,q}^{t} \left| S_i \right| =  \mathbf{v}_{p,q}^{t},
	\end{equation}
	
	\noindent since we assume cells have unit area. This essentially means that cells under a given mesh node share equally a single reservoir of virus.
	
	In the \textbf{fine mesh} case ($\dx\le1$), we write $N_i$ for the set of nodes associated with cell $i$. Assuming that $\dx = 1/k$ for some $k \in \mathbb{Z}$ is sufficient to ensure that we can position the mesh such that $N_i \cap N_j = \emptyset \quad \forall i\neq j$, as seen in the grid alignment figure in the main article. Then virus secretion is governed by
	
	\begin{equation}
	\mathbf{v}_{p,q}^{t+1} = \mathbf{v}_{p,q}^{t} + \frac{p\dt}{\dx^2} \quad \forall (p,q) \in N_i.
	\end{equation}
	
	\noindent and integration over cell domain is defined as
	
	\begin{equation}
	\int_{S_i} v(\mathbf{x},t) d\mathbf{x}  \approx  \sum_{(p,q) \in N_i}\mathbf{v}_{p,q}^{t} \dx^2.
	\end{equation}

	\section{Parameter scaling between tissue sizes} \label{app:param_scaling}

	For various reasons, it may be of interest to adapt parameters for the TIV system (or similar viral dynamics systems) between tissues of different sizes (i.e., different sizes of the cell population). However, in order to do so, the contact parameter \textbf{$\beta$}, which depends on the size of the total cell population, must be rescaled. We derive the necessary rescaling below.
	
	Recall the TIV model in fractional form as stated in the main article, where $\ntot$ is the total number of cells modelled:
	
	\begin{align} 
	\label{eq:TIV_T}
	\ddt{}\left(\frac{T}{\ntot}\right) &= -\beta \frac{T}{\ntot}V,\\
	\label{eq:TIV_I}
	\ddt{}\left(\frac{I}{\ntot}\right) &= \beta \frac{T}{\ntot}V - \delta \frac{I}{\ntot},\\
	\label{eq:TIV_V}
	\ddt{V} &= pI - cV.
	\end{align}
	
	\noindent In non-dimensional form, we have:
	
	\begin{align} \label{eq:TIV_nondim}
	\frac{d\hat{T}}{d\tau} &= -\hat{\beta} \hat{T}\hat{V},\\
	\frac{d\hat{I}}{d\tau} &= \hat{\beta} \hat{T}\hat{V} - \hat{\delta} \hat{I},\\
	\frac{d\hat{V}}{d\tau} &= \hat{I} - \hat{V}.
	\end{align}
	
	\noindent where $\hat{\beta} =\beta p \ntot/c^2 $, $\hat{\delta} = \delta/c$,  $\hat{V} = \frac{cV}{p \ntot}$, $\hat{T} = T/\ntot$, $\hat{I} = I/\ntot$, and the characteristic time scale is $L_T = 1/c$. We account for a rescaling to a smaller cell pool as follows. Say we simulate a population of $\nabm$ cells. Then we redimensionalise Equation \eqref{eq:TIV_nondim} by now defining $\tilde{T} = \nabm\hat{T}$, $\tilde{I} = \nabm\hat{I}$, $\tilde{V} = p\nabm \hat{V}/c$. The new set of equations for the smaller cell grid in dimensional quantities is as follows.
	
	\begin{align} \label{eq:TIV_rescaled}
	\ddt{}\left(\frac{\tilde{T}}{\nabm}\right) &= -\frac{\ntot}{\nabm}\beta \frac{\tilde{T}}{\nabm}\tilde{V},\\
	\ddt{}\left(\frac{\tilde{I}}{\nabm}\right) &= \frac{\ntot}{\nabm}\beta \frac{\tilde{T}}{\nabm}\tilde{V} - \delta \tilde{I},\\
	\ddt{\tilde{V}} &= p\tilde{I} - c\tilde{V}.
	\end{align}
	
	\noindent In particular, note that we rescale $\beta$ by dividing by the factor $\nabm/\ntot$, which is the proportion of the larger pool of cells which we represent in the smaller model.

	\section{Timing and spatial distribution of infections for $\rzero^*$ simulations}
	
	We analysed the distribution of infection times (for the $D=0.1~\text{CD}^2\text{h}^{-1}$ case) for different choices of coarse mesh. Subfigure (a) of Supplementary Figure \ref{fig:rzero_time_and_space} shows a histogram of infection times for the reference case, $\dx=1$, over 200 simulations, using the same set-up as in the main body of the text. Here, we stratify the data by the distance between the infection and the nearest infected cell, to give an indication of how far the virus has travelled before causing the infection. Predictably, the largest group of infections are very close to infected cells (especially neighbours), and more distant infections become more likely as time passes and virus has time to disperse. We sought to compare this distribution to the case where $\dx$ is coarse, however such a stratification becomes meaningless in this scenario. This is because, as we mentioned above, when the mesh is coarse, multiple cells share the same viral node and hence virus effectively travels instantaneously between these cells. We will term the set of cells associated with a single viral node in this scenario a "plate" of cells.
	
	In Subfigure (c) of Supplementary Figure \ref{fig:rzero_time_and_space}, we plot the distribution of infection times for $\dx=2$ and $6$ (with the same simulation set-up as above), and simplify stratify the data by whether or not infections occurred on the same plate as an infected cell. These histograms provide insight into the mechanisms driving the change in distribution for large $\dx$ values. Firstly, the acceleration in average infection time can be explained by examining the first ten hours of infections. Compared to the reference case, which has a slow uptick in infections early in the simulation, when the mesh is coarse infections begin immediately, driven by same-plate infections. This is particularly true in very coarse cases. As the distribution for $\dx=6$ shows, nearly all infection events, especially early on, occur on the same plate as infected cells. Here, the model acts as an ODE, with no spatial constraints and therefore faster infection. This explains why coarse mesh dynamics closely approximate the ODE dynamics, independently of the diffusion coefficient.
	
	By contrast, in the $\dx=2$ case, the initial burst of rapid same-plate infections appears to burn out after a fairly short period of time. We hypothesise that this deceleration and ultimate plateau of infections is due to the smaller pool of same-plate cells available, compared to very large $\dx$ cases. Once availability of this group of cells becomes constrained, we predict that the overall rate of infection will slow since cells further away from the infected cells (on neighbouring plates or further) require virus to travel \emph{further} to access compared to the $\dx=1$ case. For a demonstration, see Subfigure (b) Supplementary Figure \ref{fig:rzero_time_and_space}. In Subfigure (d) of Supplementary Figure \ref{fig:rzero_time_and_space}, we plot the proportion of the same-plate cells in the system which become infected over time. It becomes immediately clear that a far larger proportion of these cells are consumed in the smaller $\dx$ case (>80\%) compared to when $\dx$ is very large ($\sim$32\%). We thus predict that in simulations where discretisation is only slightly coarser than the reference case, the infection will experience availability constraints of same-plate cells much earlier than for very coarse discretisations. This can be seen in the curvature of both plots. The points of inflection for both plots, shown in magenta, indicate that the rate of same-plate infection begins to slow far earlier in the $\dx=2$ case compared to when $\dx=6$.
	
	It is also worth noting that this same-plate effect is felt particularly acutely when the diffusion coefficient is small, since the slow viral dispersal rate increases the time it takes for virus to reach cells at a distance from the source cell, and therefore increases the importance of same-plate infections. This may explain why the mean time of infection curves for $D=0.1~\text{CD}^2\text{h}^{-1}$ and $D=1~\text{CD}^2\text{h}^{-1}$ \emph{overlap} at coarse mesh refinement. Smaller diffusion in general causes later infections (as seen for $\dx=1/6$ through to $\dx=2$), but when the mesh is coarser than this, the presence of a large reservoir of same-plate cells eliminates spatial constraints even more effectively than the larger diffusion coefficient, and the infection distribution is skewed even further left.

	\clearpage
	\section{Supplementary figures} \label{app:SI_figures}

	\begin{figure}[h!]
		\centering
				
		\includegraphics[width=0.9\linewidth]{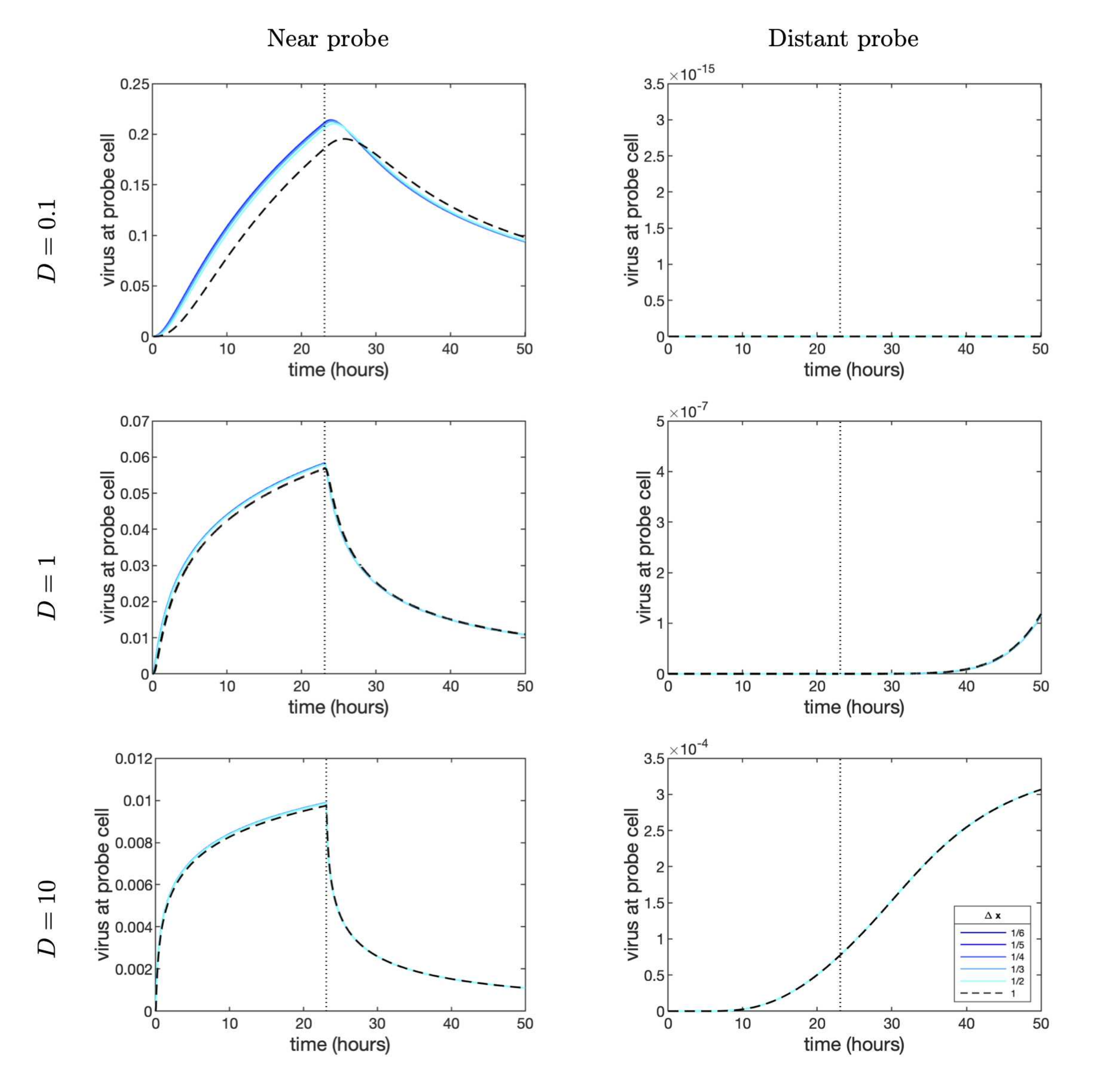}

		\caption{SUPPLEMENTARY FIGURE: Convergence of viral diffusion to cells nearby and distant from a secreting cell for fine spatial discretisation under different diffusion coefficients. Here the source is cell (30,30), the near probe is cell (31, 31), and the distant probe is cell (61, 61).}
		\label{fig:convergence_mult_diff_fine_dx}
		
	\end{figure}

	\begin{figure}
		\centering
				
		\includegraphics[width=0.9\linewidth]{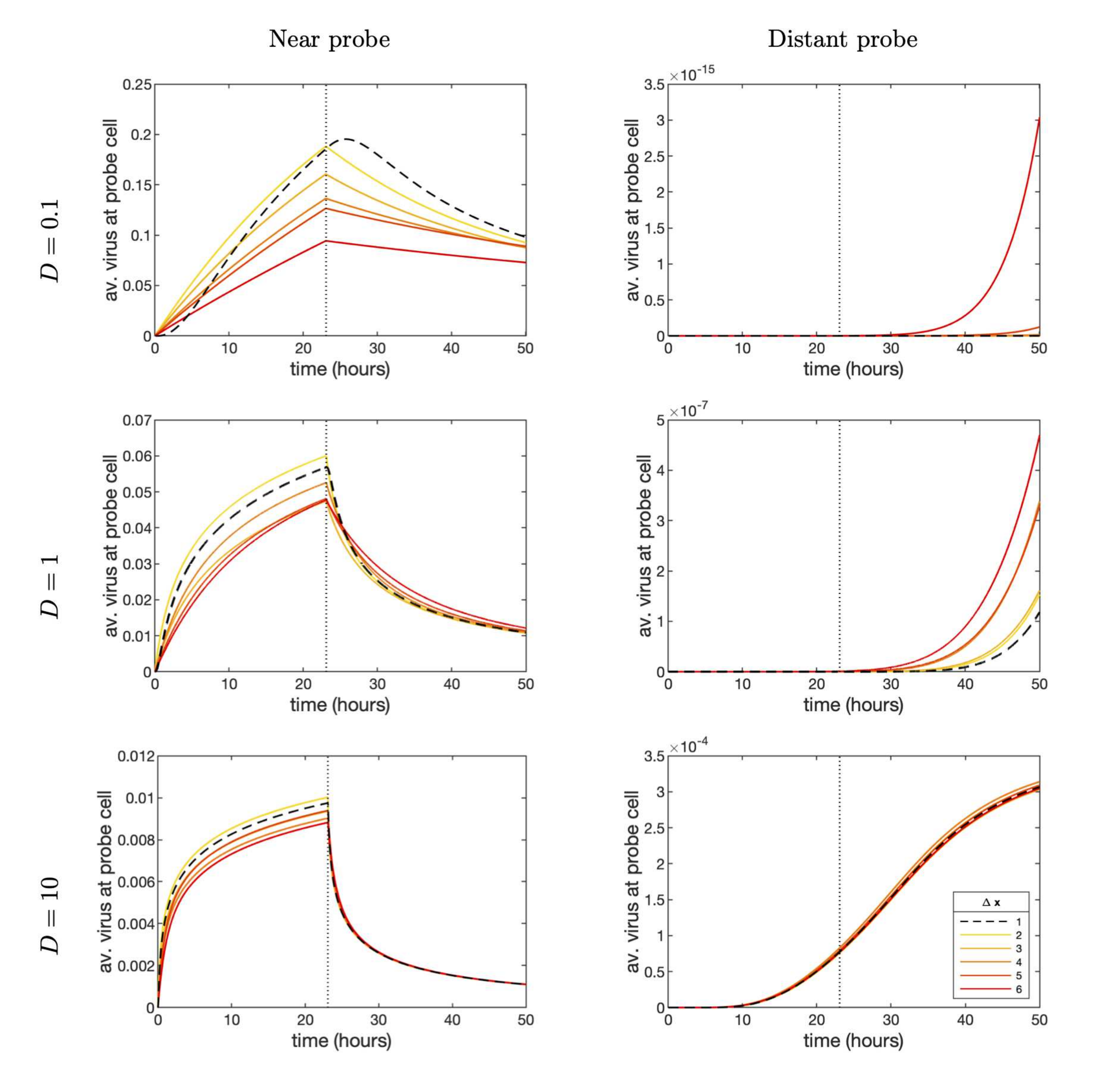}

		\caption{SUPPLEMENTARY FIGURE: Divergence of viral diffusion to cells nearby and distant from a secreting cell for coarse spatial discretisation under different diffusion coefficients. As in Figure \ref{fig:convergence_mult_diff_fine_dx}, we choose probes at a diagonal distance of 1 and 30 CD, but here we choose the probe cell at random to account for the position of the source and probe cells relative to the viral nodes. Here we display the average behaviour across 50 such simulations.}
		\label{fig:convergence_mult_diff_large_dx}
		
	\end{figure}

	\begin{figure}[h!]
		\centering

		\includegraphics[width=0.9\linewidth]{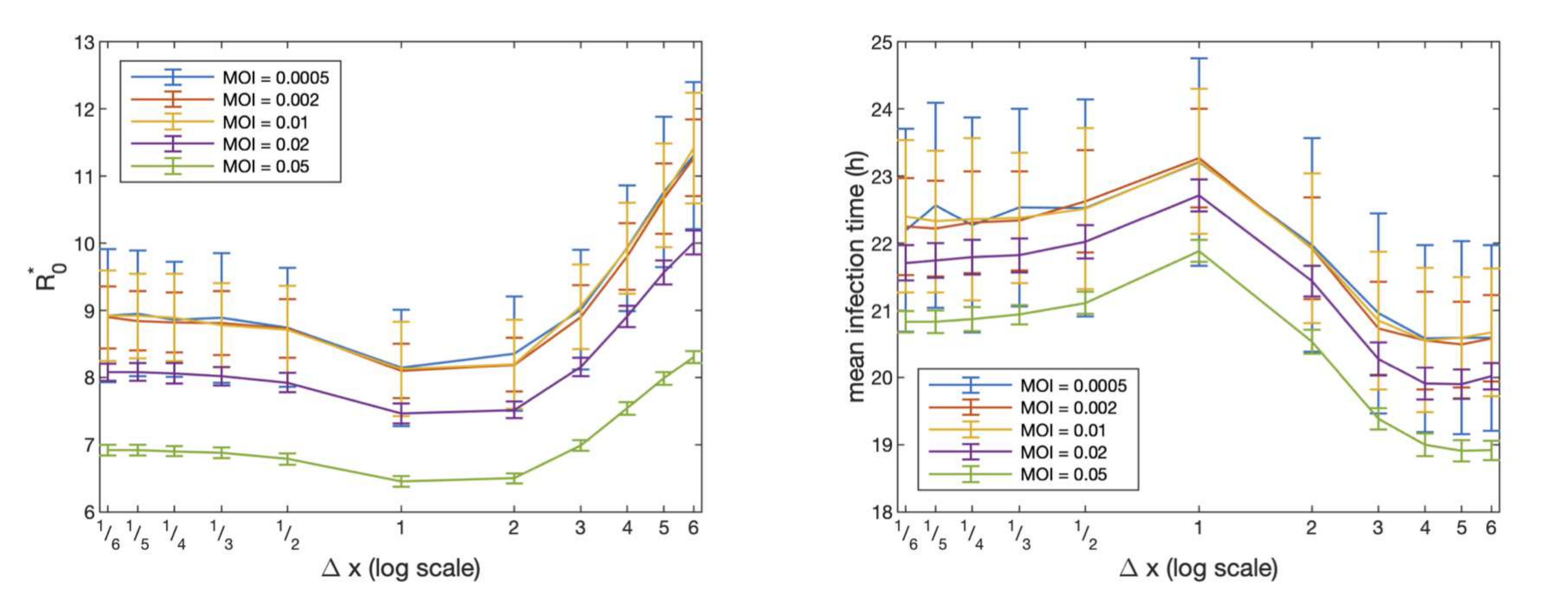}
		
		\caption{SUPPLEMENTARY FIGURE: Left, $\rzero^*$ against mesh refinement ($\dx$) for different choices of multiplicity of infection. Error bars indicate one standard deviation. Right, corresponding mean infection time. We use a 120$\times$120 grid of cells with a diffusion coefficient of $D=0.1~\text{CD}^2\text{h}^{-1}$ and take an average of 200 model simulations. Stochastic noise increases as MOI $\rightarrow 0$.}
		\label{fig:rzero_changing_moi}
	\end{figure}

	\begin{figure}[h!]
		
		\centering
				
		\includegraphics[width=0.9\linewidth]{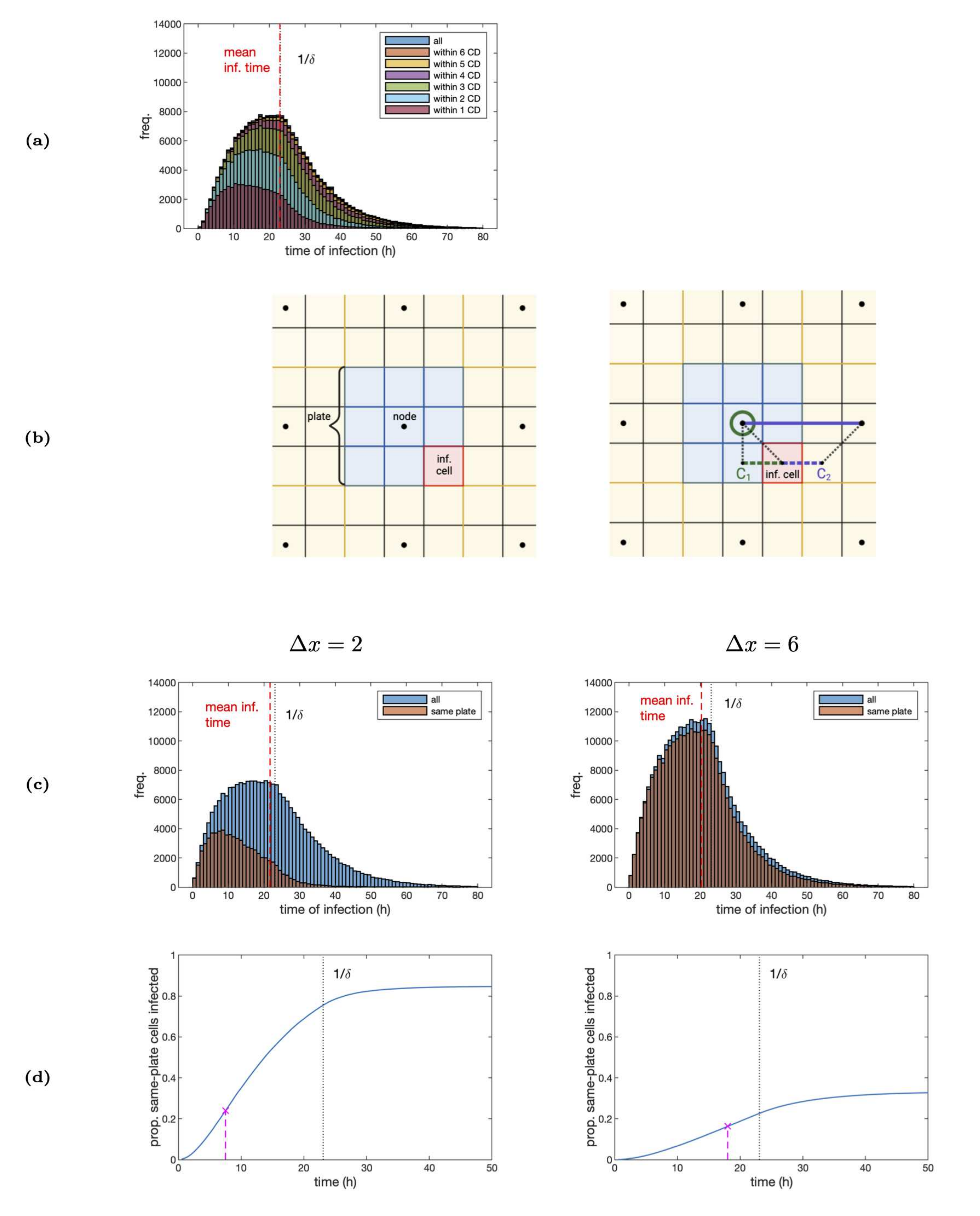}
		
		\caption{SUPPLEMENTARY FIGURE: Under coarse mesh schemes, the availability of target cells which share a viral node with infected cells drives the shape of the infection time distribution. (a) Histogram of infection times for $\dx=1$ over 200 simulations with bin width 1h, stratified by distance of infection from an infected cell in terms of cell diameters (CD). Here we use $D=0.1~\text{CD}^2\text{h}^{-1}$. (b) Left: Demonstration of a "plate" of cells under a coarse mesh scheme (here $\dx=3$). Blue cells share the same plate (same viral node) as the red infected cell. We call these "same-plate" cells. Right: Effective viral distance between the red infected cell and two cells, $C_1$ and $C_2$, under the reference scheme ($\dx=1$, dashed line) and the coarse scheme (here, $\dx=3$, solid line). (c) Histogram of infection times for $\dx=2$ and $\dx=6$ with bin width 1h, stratified by whether infections occur on the same plate as an infected cell. (d) Proportion of cells on the same plate as infected cells (same-plate cells) which are infected over time. The point of inflection is indicated in magenta.}
		\label{fig:rzero_time_and_space}
	\end{figure}

	